\documentclass{webofc}
\usepackage[varg]{txfonts}   

\usepackage{times}

\usepackage[utf8]{inputenc}

\usepackage{graphics,graphicx}
\usepackage{float}
\usepackage{amsmath, amssymb}
\usepackage{multirow}
\usepackage{longtable}
\usepackage{color}
\usepackage{slashed}
\usepackage{sidecap}
\usepackage{subfigure}

\usepackage[colorlinks=true,linktocpage=true,linkcolor=blue,citecolor=blue,hyperfootnotes=true]{hyperref}
\usepackage[normalem]{ulem}  
\usepackage{subfigure}

\begin{document}


\title{Strange matter prospects within the string-flip model}

\author{
\firstname{Niels-Uwe F.} \lastname{Bastian}\inst{1}\fnsep\thanks{\email{niels-uwe@bastian.science}}
\and
\firstname{David B.} \lastname{Blaschke}\inst{1,2,3}
\and
\firstname{Mateusz} \lastname{Cierniak}\inst{1}
\and
\firstname{Tobias} \lastname{Fischer}\inst{1}
\and
\firstname{Mark A. R.} \lastname{Kaltenborn}\inst{4,5}
\and
\firstname{Micha\l{}} \lastname{Marczenko}\inst{1}
\and
\firstname{Stefan} \lastname{Typel}\inst{6,7}
}
\institute{
Institute of Theoretical Physics, University of Wroclaw, Wroclaw, Poland
\and
Bogoliubov Laboratory for Theoretical Physics, Joint Institute for Nuclear Research, Dubna, Russia
\and
National Research Nuclear University (MEPhI), Moscow, Russia
\and
Department of Physics, The George Washington University, Washington DC, USA
\and
Los Alamos National Laboratory, Los Alamos, USA
\and
Institut f{\"u}r Kernphysik, Technische Universit{\"a}t Darmstadt, Darmstadt, Germany
\and
GSI Helmholtzzentrum f{\"u}r Schwerionenforschung, Darmstadt, Germany
}

\abstract{%
In this contribution we extend the recently developed two-flavor quark-matter string-flip model by including strange quarks. We discuss implications for compact stars.%
}


\maketitle

\section{Introduction}
\label{intro}
\begin{SCfigure}
	\hspace{2.5mm}
	\includegraphics[scale=0.275]{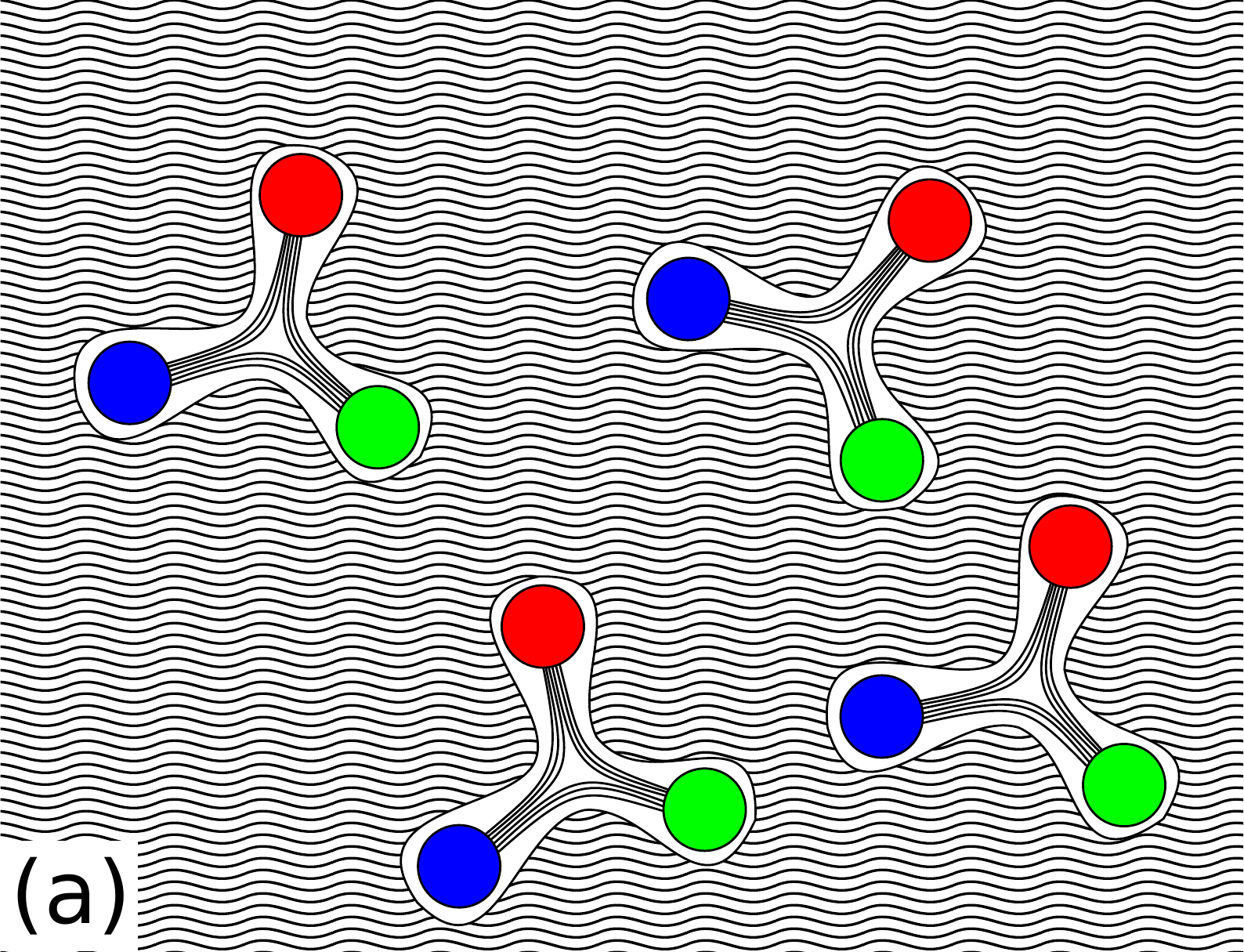}
	\includegraphics[scale=0.275]{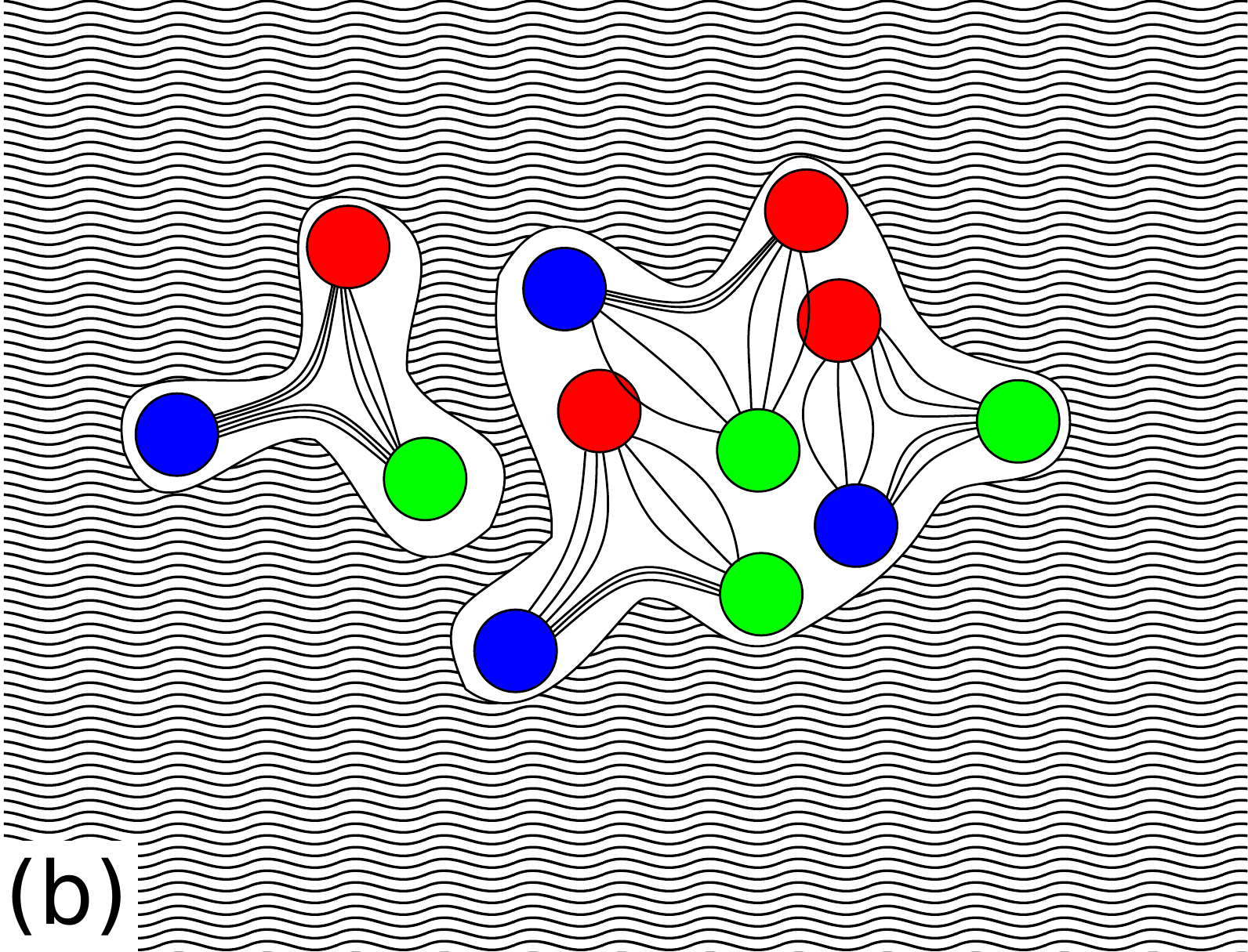}
	\caption{Illustration of the effective reduction of the string tension (density of color field lines) at high densities. At low densities (a) the field lines are compressed to thin flux tubes by the dual Meissner effect while at high densities (b) this pressure is reduced and consequently the effective string tension is lowered. Figure from \cite{Kaltenborn:2017}.}
	\label{fig1}
\end{SCfigure}

The state of matter at extreme densities encountered, e.g., in the interior of neutron stars is highly unknown.
In particular, the appearance of additional particle degrees of freedom, such as hyperons or the transition to the quark-gluon plasma, are challenged by the observations of pulsars with masses on the order of 2~M$_\odot$~\cite{Demorest:2010bx,Antoniadis:2013pzd,Fonseca:2016tux}, as they tend to generally soften the equation of state (EOS).
Despite the yet unresolved hyperon problem, it has been demonstrated recently that vector repulsion between quarks in the deconfined phase provides the necessary stiffness to yield massive neutron stars of 2~M$_\odot$ or more~\cite{Klahn:2006iw,Benic:2014jia,Klaehn:2015}.
Solutions of Quantum Chromodynamics (QCD) -- the theory of strong interactions among quarks and gluons -- at vanishing chemical potential and finite temperature predict a smooth crossover transition at $T\simeq 154\pm9$~MeV~\cite{Katz:2014PhLB} from hadrons to quarks and gluons.
For astrophysical studies at low temperatures and large densities, only phenomenological approaches are available, such as, the thermodynamic bag model~\cite{Farhi:1984qu} and models based on the Nambu-Jona-Lasinio (NJL) approach~\cite{Nambu:1961tp,Klevansky:1992qe,Buballa:2003qv}.
These models feature a first-order hadron-quark phase transition for a given hadronic EOS and take (de)confinement into account effectively.
Note that perturbative QCD is valid only at extremely high densities and temperatures~\cite{Kurkela:2014vha}, when quarks and gluons are no longer strongly interacting.
This goes far beyond the conditions encountered in astrophysical applications, e.g., neutron stars and supernovae.
Recently, a new approach has been developed, which starts from an effective relativistic density-functional~\cite{Kaltenborn:2017} that implements quark confinement based on the string-flip model~(SFM)~\cite{Horowitz:1985tx,Ropke:1986qs}.
According to the SFM, the color interactions between quarks are saturated among nearest neighbors.
In hadronic matter they are confined to the same hadron, while in deconfined matter the situation is described by a string-length distribution and an effective string tension as sketched in Fig.~\ref{fig1}. 

\section{String-flip model for two-flavor quark matter}

\begin{figure}[t!]
	\subfigure{\includegraphics[scale=0.28]{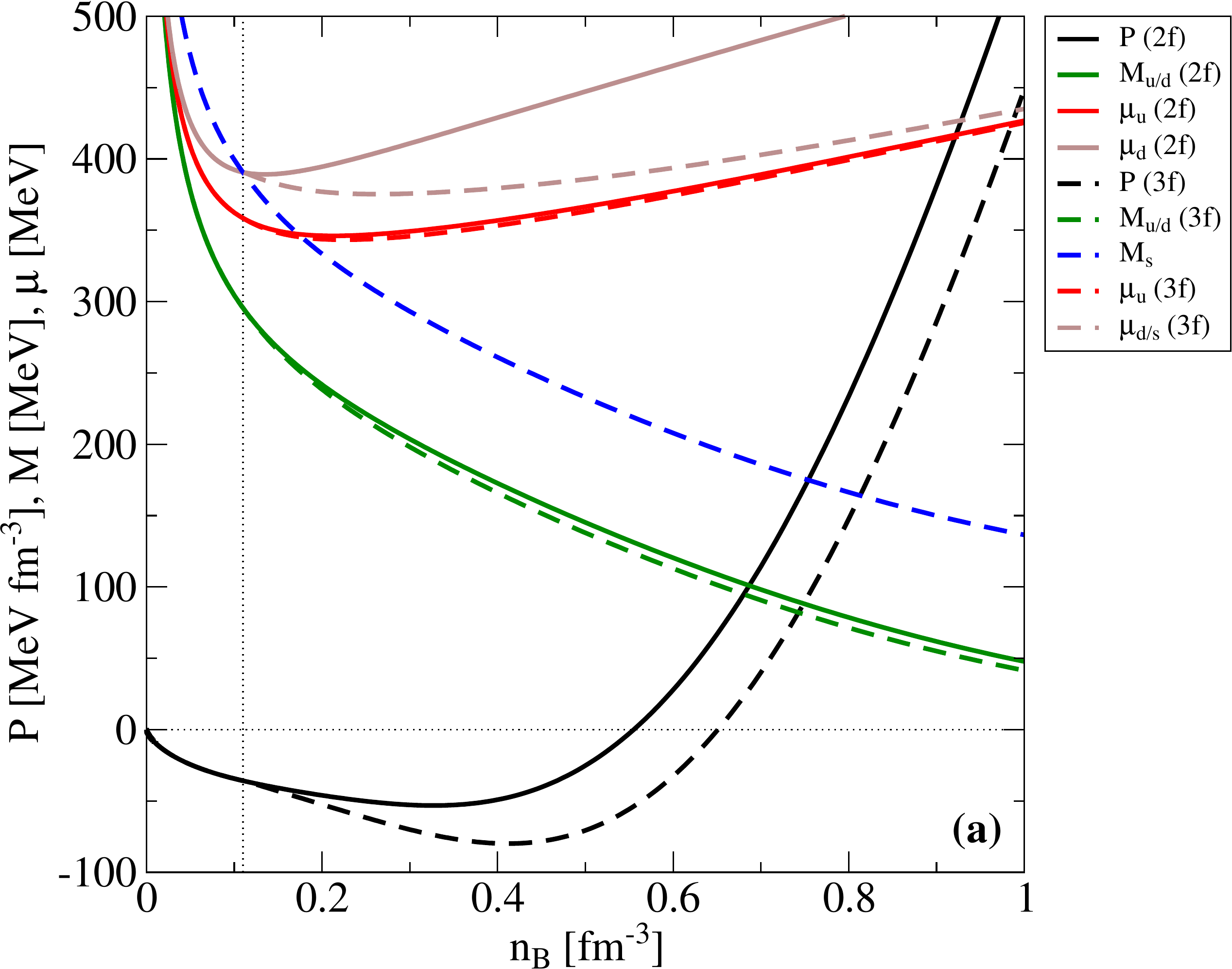}\label{fig2a}}
	\hfill
	\subfigure{\includegraphics[scale=0.28]{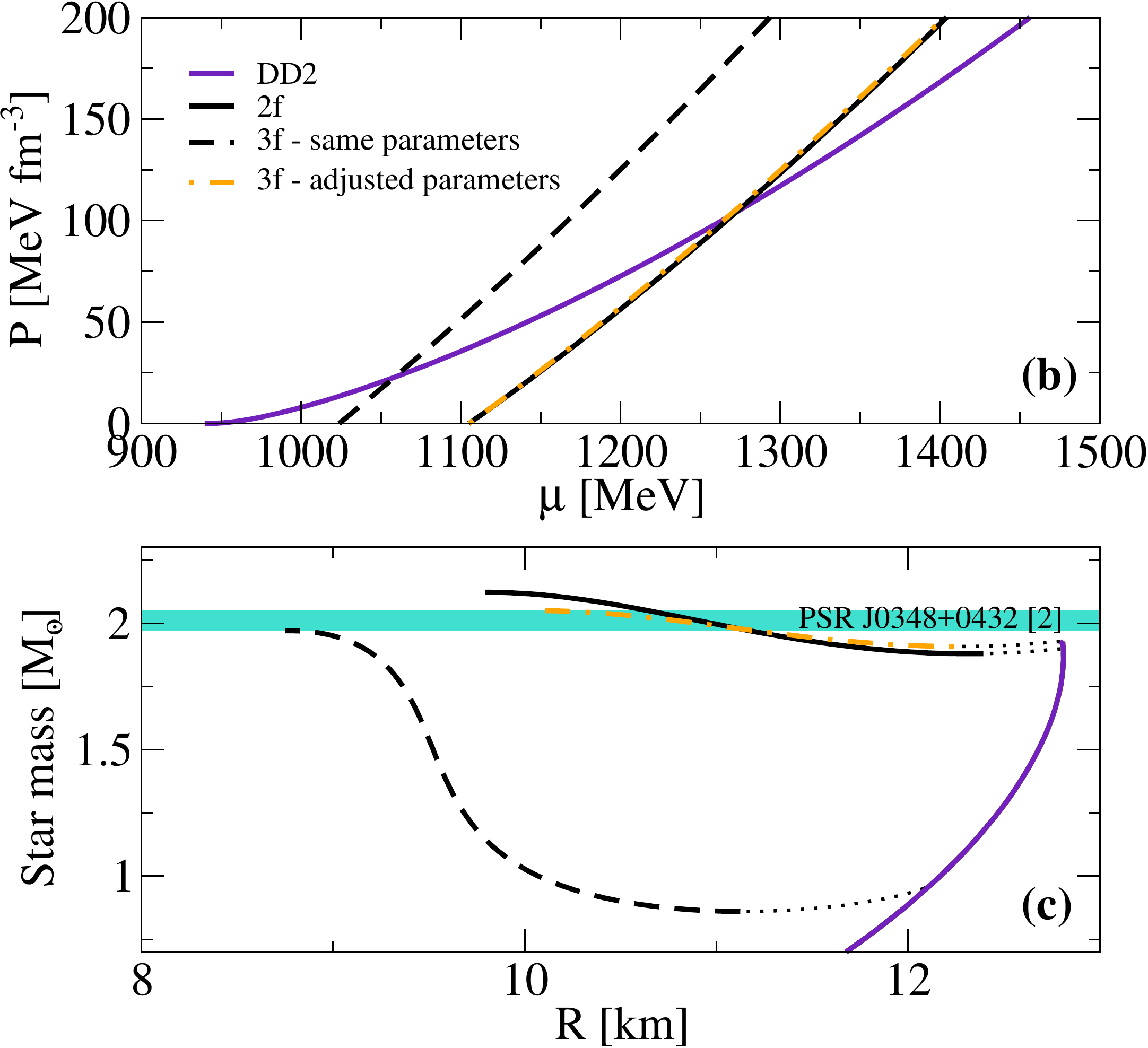}\label{fig2b}}
	\caption{%
		$\beta$-equilibrium SFM EOS for two- and three-flavor quark matter, as well as corresponding hybrid-star configurations, for the parameters $\alpha = 0.2$~fm$^6$, $\sqrt{D_0} = 240$~MeV, $a = -2.0$~MeV~fm$^3$, $b=2.0$~MeV~fm$^9$, and $c=0.036$~fm$^6$. %
		(a) Effective quark masses, effective chemical potentials and pressure. The vertical dotted line shows the onset of strange quarks. %
		(b) Maxwell construction of the phase transition between the DD2 hadronic and three quark matter EOS. %
		(c) Mass-radius relation for the three resulting hybrid EOS: dotted lines indicate instability of the configuration against gravitational collapse.%
		}
	\label{fig2}
\end{figure}

In the mean-field approximation, the correlation energy can be obtained by folding the string-length distribution function for a given density with some interaction potential~\cite{Horowitz:1991fn,Ropke:1986qs,Horowitz:1991ux}. Moreover, the average string length between quarks in uniform matter is related to the scalar density, $n_\mathrm s$, being proportional to ${n_\mathrm s^{-1/3}}$. Hence, one obtains a contribution to the energy density functional of quark matter correspondingly~\cite{Khvorostukin:2006aw}. In analogy to the Walecka model of nuclear matter~\cite{Kapusta:1989tk}, the relativistic density-functional approach to interacting quark matter can be obtained from the path integral approach based on the partition function \cite{Kaltenborn:2017}, 
\begin{align}
	\label{Z}
	\mathcal{Z} &= \int \mathcal{D}\bar{q}\mathcal{D}q \exp\left\{\int_0^\beta d\tau\int_Vd^3 x \left[\mathcal{L}_{\rm eff} + \bar{q}\gamma_0 \hat{\mu} q\right]\right\}~,\;\;\;\rm with\;\;q=\left(\begin{array}{c}q_u\\q_d\end{array}\right)~,
\end{align}
with effective Lagrangian density $\mathcal{L}_{\textrm{eff}}=\mathcal{L}_{\textrm{free}}- U(\bar{q}q, \bar{q}\gamma_0q)$. The interaction is given by the potential $U(\bar{q}q, \bar{q}\gamma_0q)$, which is a nonlinear functional of the scalar and vector quark field-currents.
In the mean-field approximation, this potential can be expanded around the expectation values of the field currents, $n_{\rm s}$ and $n_{\rm v}$ respectively,
\begin{align}
	U(\bar{q}q,\, \bar{q}\gamma_0q) &= U(n_{\rm s}, n_{\rm v}) + (\bar{q}q -  n_{\rm s})\Sigma_\mathrm {s} + (\bar{q}\gamma_0q -  n_{\rm v}) \Sigma_\mathrm {v} +\ldots~,
\end{align}
with scalar and vector self-energies, $\Sigma_{\rm s}$ and $\Sigma_{\rm v}$. Moreover, the following density functional of the interaction is adopted,
\begin{align}
	\label{eq:potential}
	U({n}_\mathrm s,{n}_\mathrm v) &= D({n}_\mathrm v){n}_\mathrm s^{2/3} +a {n}_\mathrm v^2 + \frac{b {n}_\mathrm v^4}{1+c {n}_\mathrm v^2}~.
\end{align}
The first term captures aspects of (quark) confinement through the density dependent scalar self-energy, $\Sigma_\mathrm s = \frac{2}{3}D({n}_\mathrm v){n}_\mathrm s^{-1/3}$, defining the effective quark mass $M = m + \Sigma_\mathrm s$ (see Fig.~\ref{fig2}).
We also employ higher-order quark interactions~\cite{Klahn:2006iw}, by inclusion of the third term in Eq.\eqref{eq:potential}, for the description of hybrid stars (neutron stars with a quark matter core) in order to obey the observational constraint of $2~$M$_\odot$.
To this end, the denominator in the last term of Eq.\eqref{eq:potential} guarantees that for the appropriate choice of the parameters $b$ and $c$, causality is not violated (i.e., the speed of sound $c_\mathrm s=\sqrt{\partial P/\partial \varepsilon}$ does not exceed the speed of light).
All terms in Eq.\eqref{eq:potential} that contain the vector density contribute to the shift defining the effective chemical potentials $\mu^* = \mu - \Sigma_\mathrm V$, $\Sigma_{\rm v} = 2an_{\rm v} + \frac{4b n_{\rm v}^3}{1+c n_{\rm v}^2} - \frac{2 b c n_{\rm v}^5}{(1+c n_{\rm v}^2)^2} + \frac{\partial D(n_{\rm v})}{\partial n_{\rm v}}{n}_s^{2/3}$.
The SFM modification takes into account the effective reduction of the in-medium string tension, $D(n_\mathrm v) = D_0 \Phi(n_{\rm v})$.
It is understood as a consequence of the modification of the pressure on the color field lines by the dual Meissner effect, since the reduction of the available volume corresponds to the reduction of the non-perturbative dual superconductor QCD vacuum that determines the strength of the confining potential between the quarks.
The reduction of the string tension is modeled via a Gaussian functional, $\Phi(n_{\rm v})=e^{- \alpha n_{\rm v}^2}$,  where $\alpha$ is a parameter for the available volume fraction.
A detailed discussion of the role of the parameters $a,b,c$ and $\alpha$ is given in Ref.~\cite{Kaltenborn:2017}.

\section{Extension to strange matter -- implications for compact stars}
In order to extend the SFM to three-flavors, we extend quark spinor field, $(q_{\rm u}, q_{\rm d}) \longrightarrow (q_{\rm u}, q_{\rm d}, q_{\rm s})$.
Note that flavor-dependent interactions are neglected at this stage, and, hence, all quark flavors couple and contribute to the mean fields equally.
Consequently, the total scalar and vector densities of quarks are given by the following expressions, $n_\mathrm v = n_\mathrm v^u +  n_\mathrm v^d +  n_\mathrm v^s$ and $n_\mathrm s = n_\mathrm s^u +  n_\mathrm s^d +  n_\mathrm s^s$, respectively.
We are interested in applications of the SFM to compact stars, where matter is in the state of $\beta$-equilibrium at zero temperature via weak processes, resulting in the relation between the chemical potentials for electrons ($\mu_\mathrm e$), up ($\mu_{\rm u}$), down ($\mu_{\rm d}$) and strange ($\mu_{\rm s}$) quarks, so that $\mu_\mathrm s = \mu_{\rm d} = \mu_\mathrm e + \mu_{\rm u}$.
However, strange quarks have a significantly higher bare mass compared to the light up and down quarks.
Here we assume $m_{\rm s} = 100$~MeV and $m_{\rm u}=m_{\rm d} = 5$~MeV.
The density dependence of the effective chemical potentials and effective masses is illustrated in Fig.~\ref{fig2}~(a).
This aspect is similar to chiral quark models like the NJL.
The difference is in the behaviour at the onset.
Up to the density where $\mu_{\rm s} = m_{\rm s}$, two- and three-flavor quark matter are indistinguishable.
Beyond this point (vertical dashed line in Fig.~\ref{fig2}~(a)) up- and down-quark masses drop significantly faster with density.
Note further that since strange and down quarks carry the same negative charge, their chemical potentials are somewhat lower in the three-flavor case than the down quark chemical potential in the two-flavor case. 

Since we work in the mean-field approximation, the SFM cannot provide the EOS of the confined phase.
For the hadronic EOS we select the relativistic mean-field EOS with density-dependent couplings (DD2) from Ref.~\cite{Typel:2009sy} (Fig.~\ref{fig2}~(b)).
The hadron-quark phase transition is obtained by a Maxwell construction.
SFM parameters (given in the caption of Fig.~\ref{fig2}) are selected such that massive hybrid star configurations are obtained with maximum masses $\gtrsim 2$~M$_\odot$ (see Fig.~\ref{fig2}~(c)), featuring here the `twin-phenomenon'~\cite{Bhattacharyya:2004fn,Glendenning:1998ag} where stable hadronic and hybrid star branches are disconnected by an unstable branch~\cite{Alford:2013aca} (thin black dotted lines in Fig~\ref{fig2}~(c)).
It has recently been shown \cite{Alvarez-Castillo:2017qki} that this feature can also be obtained within the multi-polytrope approach \cite{Hebeler:2013nza}.
The appearance of strange quarks softens the EOS significantly, which results in the shift of the phase-transition onset to somewhat lower densities for the same SFM parameters (see Fig~\ref{fig2}~(b)) and a lower onset mass (see Fig~\ref{fig2}~(c)).
This is compensated when re-adjusting the parameter set for the three-flavor case accordingly ($\alpha = 0.14$~fm$^6$, $\sqrt{D_0} = 240$~MeV, $a = -1.0$~MeV~fm$^3$, $b=4.0$~MeV~fm$^9$, $c=0.036$~fm$^6$).

\section{Summary}
Consequences of strangeness in quark matter for compact stars are studied within the SFM, a new phenomenological approach to quark-matter with an effective confining mean field.
We observe an onset of strange quarks at much lower densities than the deconfinement phase transition.
This leads to the simultaneous appearance of all flavours, in contrast to the commonly employed NJL-type models with sequential deconfinement~\cite{Klaehn:2015,Blaschke:2008br}.
We have found that the inclusion of the strange quark flavor lowers the onset of the deconfinement transition, but SFM parameters can be selected such that hybrid star sequences in the mass-radius diagram of two- and three-flavor quark matter are nearly indistinguishable.

\bigskip
{\bf Acknowledgments} This work was supported by the Polish National Science Center (NCN) under grant numbers UMO-2014/13/B/ST9/02621 (NUFB, MC, MM and MARK), UMO-2016/23/B/ST2/00720~(TF) and UMO-2011/02/A/ST2/00306 (DBB). DBB acknowledges support by the MEPhI Academic Excellence Project under contract No.~02.a03.21.0005. ST was partially supported by the DFG through grant SFB1245. This work was supported in part by the COST Actions MP1304 ``NewCompStar", CA15213 ``THOR'' and CA16117 ``ChETEC''.


\begin{thebibliography}{}
\footnotesize


\bibitem{Demorest:2010bx}
P.~Demorest, T.~Pennucci, S.~Ransom, M.~Roberts and J.~Hessels,
Nature {\bf 467}, 1081 (2010).

\bibitem{Antoniadis:2013pzd}
J.~Antoniadis {\it et al.},
Science {\bf 340}, 6131 (2013).

\bibitem{Fonseca:2016tux} 
  E.~Fonseca {\it et al.},
  Astrophys.\ J.\  {\bf 832}, no. 2, 167 (2016).


\bibitem{Klahn:2006iw}
  T.~Kl{\"a}hn, \textit{et.al.},
  Phys.\ Lett.\ B {\bf 654}, 170 (2007).

\bibitem{Klaehn:2015}
T.~Kl{\"a}hn and T.~Fischer,
Astrophs.\ J. {\bf 810}, 134 (2015).

\bibitem{Benic:2014jia}
S.~Benic, D.~Blaschke, D.~E.~Alvarez-Castillo, T.~Fischer and S.~Typel,
Astron.\ Astrophys.\  {\bf 577}, A40 (2015).

\bibitem{Katz:2014PhLB}
S.~Bors{\'a}nyi, Z.~Fodor, C.~Hoelbling, S.~D.~Katz, S.~Krieg, and K.~K.~Szab{\'o},
Phys.\ Lett.\ B {\bf 730}, 99 (2014).

\bibitem{Farhi:1984qu}
E.~Farhi and R.~Jaffe,
Phys.\ Rev.\ D {\bf 30}, 2379 (1984).

\bibitem{Nambu:1961tp}
Y.~Nambu anf G.~Jona-Lasinio,
Phys.\ Rev. {\bf 122}, 345 (1961).

\bibitem{Klevansky:1992qe}
S.~Klevansky,
Rev.\ Mod.\ Phys. {\bf 64}, 649 (1992).

\bibitem{Buballa:2003qv}
M.~Buballa,
Phys.\ Rept. {\bf 407}, 205 (2005).

\bibitem{Kurkela:2014vha}
A.~Kurkela, E.~S.~Fraga, J.~Schaffner-Bielich, A.~Vuorinen,
Astrophys.\ J. {\bf 789}, 127 (2014).

\bibitem{Kaltenborn:2017}
M.~A.~R.~Kaltenborn, N.~U.~F.~Bastian and D.~B.~Blaschke,
Phys.\ Rev.\ D {\bf 96}, 056024 (2017).

\bibitem{Horowitz:1985tx}
C.~J.~Horowitz, E.~J.~Moniz and J.~W.~Negele,
Phys.\ Rev.\ D {\bf 31}, 1689 (1985).

\bibitem{Ropke:1986qs}
G.~R\"opke, D.~Blaschke and H.~Schulz,
Phys.\ Rev.\ D {\bf 34}, 3499 (1986).

\bibitem{Horowitz:1991fn}
C.~J.~Horowitz and J.~Piekarewicz,
Nucl.\ Phys.\ A {\bf 536}, 669 (1992).

\bibitem{Horowitz:1991ux}
C.~J.~Horowitz and J.~Piekarewicz,
Phys.\ Rev.\ C {\bf 44}, 2753 (1991).

\bibitem{Khvorostukin:2006aw}
A.~S.~Khvorostukin, V.~V.~Skokov, V.~D.~Toneev and K.~Redlich,
Eur.\ Phys.\ J.\ C {\bf 48}, 531 (2006).

\bibitem{Kapusta:1989tk}
J.~I.~Kapusta,
``Finite Temperature Field Theory'',
Cambridge University Press, Cambridge (1989).

\bibitem{Typel:2009sy}
S.~Typel, G.~R\"opke, T.~Kl\"ahn, D.~Blaschke and H.~H.~Wolter,
Phys.\ Rev.\ C {\bf 81}, 015803 (2010).

\bibitem{Glendenning:1998ag}
  N.~K.~Glendenning and C.~Kettner,
  Astron.\ Astrophys.\  {\bf 353}, L9 (2000).

\bibitem{Bhattacharyya:2004fn}
  A.~Bhattacharyya, S.~K.~Ghosh, M.~Hanauske and S.~Raha,
  Astron.\ Astrophys.\  {\bf 418}, 795 (2004).

\bibitem{Alford:2013aca}
M.~G.~Alford, S.~Han and M.~Prakash,
Phys.\ Rev.\ D {\bf 88}, no. 8, 083013 (2013).

\bibitem{Alvarez-Castillo:2017qki}
  D.~E.~Alvarez-Castillo and D.~B.~Blaschke,
  Phys. Rev. C 96, 045809 (2017)

\bibitem{Hebeler:2013nza}
K.~Hebeler, J.~M.~Lattimer, C.~J.~Pethick and A.~Schwenk,
Astrophys.\ J.\  {\bf 773}, 11 (2013).

\bibitem{Blaschke:2008br}
  D.~Blaschke, F.~Sandin, T.~Klahn and J.~Berdermann,
  Phys.\ Rev.\ C {\bf 80} (2009) 065807

\end{thebibliography}
\end{document}